\documentclass[showpacs,prl,a4,twocolumn]{revtex4}
\usepackage{graphicx}
\usepackage[dvips]{color}

\newcommand{\av}[1]{ {\left\langle #1 \right\rangle} }

\newcommand{\rd}{ {\rm{d}} }

\newcommand{\trd}[1]{ #1 }

\begin{document}

\title[Amplitude Noise Suppression in Cavity-Driven Oscillations]{Amplitude Noise Supression in Cavity-Driven Oscillations of a Mechanical Resonator}%
\author{D. A. Rodrigues 
 and A. D. Armour}
\address{School of Physics and Astronomy, University of
Nottingham, Nottingham NG7 2RD, U.K.}%


\begin{abstract}
We analyze the amplitude and phase noise of limit-cycle oscillations in
a mechanical resonator coupled parametrically to an optical cavity driven above its resonant frequency.
At a given temperature the limit-cycle oscillations
 have lower amplitude noise than states of the same average amplitude excited by a pure harmonic drive;
for sufficiently low thermal noise a sub-Poissonian resonator state
can be produced. We also calculate the linewidth narrowing that
occurs in the limit-cycle states and show that while the minimum is
set by direct phase diffusion, diffusion due to the optical spring
effect can dominate if the cavity is not driven exactly at a
side-band resonance.
\end{abstract}

\pacs{42.50.Lc, 42.50.Pq, 07.10.Cm}
\maketitle

Recently there has been considerable progress towards the goal
of observing signatures of quantum behavior in the collective
vibrations of mechanical resonators\,\cite{AS}. Evidence of quantum behavior
is expected to be found in the production of non-classical states
of mechanical motion such as squeezed states or superpositions of spatially separated states\,\cite{mpb,AS}.
Quantum effects should also be evident in the non-linear dynamics of mechanical resonators\,\cite{ron}, as well as
in the fundamental limits on the sensitivity with which mechanical motion can
be monitored\,\cite{limits}. However, because mechanical resonators typically have resonant frequencies in the
radio-frequency range or below, thermal fluctuations would naturally
tend to mask the quantum features and hence significant efforts have been devoted
to developing ways of cooling a mechanical resonator down to its
ground state\,\cite{AS,omechrev}.

The radiation pressure force which arises when a mechanical resonator is coupled parametrically
to a driven optical cavity\,\cite{omechrev} provides a
very effective way of suppressing thermal fluctuations in mechanical resonators. When the cavity is driven
below resonance, quanta are absorbed from the mechanical resonator by the cavity.
The low level of photon noise in the cavity means that, provided the relaxation rate of the cavity is much less than the mechanical frequency (the good cavity limit), the mechanical resonator can in principle be cooled almost all the
way to its ground state\,\cite{MCCG}. However, in practice the cooling effect of the cavity competes with the resonator's thermal environment
and recent experiments\,\cite{recent} have combined the driven cavity with cryogenic cooling to achieve lower occupation numbers.

If the cavity is instead driven above resonance then energy is
absorbed by the mechanical resonator leading to states of
self-sustaining oscillation\,\cite{LC,MHG}. Increasing the power of
the cavity drive leads eventually to a region of multistability
marked by a sequence of dynamical transitions between limit-cycle
states of different sizes\,\cite{LC,MHG}. Little attention has so
far been devoted to studying the quantum aspects of the limit-cycle
dynamics, although recent numerical calculations began to explore
the behavior in this regime\,\cite{LKM}. However, similar laser-like
states have been studied in mechanical oscillators coupled to a
range of finite-level systems\,\cite{CB,HRA,vpl}.

 In this Letter, we
present an analytic calculation of the amplitude noise of a
cavity-driven mechanical resonator within a limit-cycle. Our
principal finding is that the amplitude noise in a limit-cycle can
be very low: for very low thermal noise the resonator can be driven
into a sub-Poissonian state by the cavity. More generally, at a
given temperature the amplitude noise in a limit-cycle state can be
substantially lower than in an equivalent one produced by a perfect
harmonic drive. We also explore the behavior of the resonator
linewidth in the limit-cycle state, generalizing a previous
calculation\,\cite{vah}.

The parametrically coupled driven cavity and mechanical resonator system is described by\,\cite{MCCG,MHG},
\begin{equation}
H= -\hbar[\Delta+\frac{g}{2}(b+b^\dag)]a^\dag a +\hbar\omega_m
b^\dag b + \hbar\Omega(a+a^\dag),
\end{equation}
where $g$ is the coupling strength, $a$($b$) is a cavity (resonator) lowering operator,
  $\omega_m$ is the mechanical frequency and $\Omega$ parameterizes the strength of the laser drive.
   The cavity is driven at a frequency $\omega_d$ detuned from the cavity frequency, $\omega_c$ ($\gg \omega_m$),
    by $\Delta=\omega_d-\omega_c$. The
evolution of the system is described by the master
equation\,\cite{LKM},
\begin{equation}
\dot{\rho}=-\frac{i}{\hbar}[H,\rho]+\mathcal{L}_m\rho+\mathcal{L}_c\rho, \label{eq:master}
\end{equation}
where the coupling of the mechanical resonator to its thermalized
surroundings at temperature $T$ is described by,
\begin{eqnarray*}
\mathcal{L}_{m}\rho&=&-\frac{\gamma_{m}}{2}(\overline{n}+1)\left(
b^{\dagger}b\rho+\rho b^{\dagger}b-2b\rho b^{\dagger}\right)
 \\ &&-\frac{\gamma_{m}}{2}\overline{n}\left(
bb^{\dagger}\rho+\rho bb^{\dagger}-2b^{\dagger}\rho b\right),
\nonumber
\end{eqnarray*}
with $\gamma_m$ the mechanical damping rate and $\bar{n}=[{\rm exp}(\hbar\omega_m/k_{\rm B}T)-1]^{-1}$. The cavity dissipation is described by,
$\mathcal{L}_{c}\rho=-\gamma_{c}(a^{\dagger}a\rho+\rho a^{\dagger}a-2a\rho a^{\dagger})/2$
with $\gamma_c$ the decay rate and we assume $\hbar\omega_c\gg k_{\rm B}T$ so that we can neglect thermal fluctuations.

We proceed by carrying out a Wigner transformation of the master equation, which introduces the complex variables $\alpha$ and $\beta$ for the phase space of the cavity and resonator respectively\,\cite{WM}. Neglecting third-order derivative terms (truncated Wigner function approximation) in the resulting equation of motion for the Wigner function leads to a standard Fokker-Planck equation from which we obtain
the coupled Langevin equations,
\begin{eqnarray}
\dot \alpha &=& i[\Delta +\frac{g}{2}(\beta+\beta^*)]\alpha-i\Omega
-\frac{\gamma_c}{2}
\alpha+\eta_\alpha\label{eq:LVa}\\
\dot{\beta} &=& i \frac{g}{2}
\left(\alpha^*\alpha-\frac{1}{2}\right) - i\omega_m \beta
-\frac{\gamma_m}{2}\beta+\eta_\beta. \label{eq:LVb}
\end{eqnarray}
The stochastic force terms\,\cite{WM} $\eta_{\alpha},\eta_{\beta}$
have zero means and non-zero second order moments
$\av{\eta_{\alpha^*}(t)\eta_{\alpha}(t')}=\delta(t-t')\gamma_c/2$
and
$\av{\eta_{\beta^*}(t)\eta_{\beta}(t')}=\delta(t-t')\gamma_m(\bar{n}+\frac{1}{2})$.
The truncated Wigner function approximation is expected to describe
small linear fluctuations\,\cite{WM} and hence should provide a good
description for the limit-cycle states.

We follow the approach used by Marquardt et al.\,\cite{MHG} to solve
the corresponding classical dynamics and extend this to include the
noise. We make the realistic assumption that the total resonator
damping is much lower than the cavity decay rate so that the
amplitude and phase of the resonator change only very slowly on the
time-scale of the cavity dynamics. The problem is now split into two
parts. First, we solve for $\alpha$ using the ansatz
$\beta(t)=\beta_c+B{\rm e}^{-i\phi}{\rm e}^{-i\omega_mt}$. The
resulting solution is split into average and fluctuating parts,
$\alpha(t)=\langle \alpha(t)\rangle+\delta\alpha$ (where the average
corresponds to the solution obtained when the stochastic force term
is dropped) and then assuming weak fluctuations we approximate
$\alpha\alpha^*\simeq\langle \alpha
\alpha^*\rangle+\delta\alpha^*\langle
\alpha\rangle+\delta\alpha\langle \alpha^*\rangle$ to obtain an
effective equation of motion for
$\tilde{\beta}(t)=\beta(t)-\beta_c$.

Solving for the cavity dynamics and taking the Fourier transform, we obtain
\begin{equation}
\langle
\alpha'(\omega)\rangle=\sum_n\alpha_n\delta(\omega-\omega_mn)=
\frac{-i\Omega\sum_nJ_{-n}(z){\rm e}^{i\phi
n}\delta(\omega-\omega_mn)}{\gamma_c/2+i(\omega-\tilde\Delta)}\nonumber
\end{equation}
and
$\delta\alpha'(\omega)={\eta'_{\alpha}}/[{\gamma_c/2+i(\omega-\tilde\Delta)}]$,
where $z=gB/\omega_m$, $J_n(z)$ is a Bessel function of the first
kind, $\tilde\Delta=\Delta+g{\rm Re}[\beta_c]$ and the primes denote
e.g.\ $\eta'_{\alpha}(t)=\eta_{\alpha}(t){\rm e}^{-iz\sin(\phi+\omega
t)}$. Keeping only the fundamental oscillating component of $\langle
\alpha\rangle\langle \alpha^*\rangle$, we obtain
\begin{equation}
\dot{\tilde{\beta}}=-\frac{(\gamma_{BA}+\gamma_m)}{2}\tilde{\beta}-i(\omega_m+\delta\omega_m)\tilde{\beta}+\eta_{\beta}
+\frac{ig}{2}(\langle\alpha^*\rangle\delta\alpha+\langle\alpha\rangle\delta\alpha^*),\label{eq:betatilde}
\end{equation}
where the effective damping and
frequency shift of the resonator due to the cavity are given by\,\cite{MHG},
\begin{eqnarray}
\frac{\gamma_{BA}}{2}+i\delta\omega_m &=&  \frac{-ig\Omega^2}{2B}
\sum\limits_{n} \frac{ J_{n}(z)J_{n+1}(z)}{h_{n} h_{n+1}^*},
\label{eq:gamma}
\end{eqnarray}
with $h_n=\frac{\gamma_c}{2}+i(\tilde\Delta+n\omega_m )$. The center
of the mechanical oscillations is given by,
\begin{eqnarray}
\beta_c &=& \frac{ig}{2}\frac{ \sum\limits_n|\alpha_n|^2}{i\omega_m
+\frac{\gamma_m}{2}} \label{eq:bconst},
\end{eqnarray}
which, although non-linear, is well approximated by its linear form
for weak $g$.

We focus for now on the fluctuations in the amplitude of the
resonator motion. The equation of motion for the amplitude can be
written as,
\begin{equation}
\dot{B}=-\frac{\gamma_T}{2}B +\eta_{T}^-, \label{eq:Beom}
\end{equation}
where $\gamma_T=\gamma_m+\gamma_{BA}$ and  $\eta_{T}^\mp =
\frac{1}{2}(\eta_{\beta} {\rm e}^{i(\phi+ \omega_m
t)}\pm\eta_{\beta^*} {\rm e}^{-i(\phi+ \omega_m t)})
+\frac{ig}{4}(\av{\alpha^*} \delta \alpha+\av{\alpha} \delta
\alpha^*)( {\rm e}^{i (\phi+\omega_m t)}\mp {\rm e}^{-i
(\phi+\omega_m t)})$. The term $\eta_T^+$ relates to the phase
diffusion, as discussed below.

An effective diffusion constant valid on timescales long compared to
$\omega_m$ and $\gamma_c$ is obtained\,\cite{Lax} from the
zero-frequency component of the correlator $\langle
\eta_{T}^-(t)\eta_{T}^-(t')\rangle$, averaged over a mechanical period
to eliminate explicit time dependence,
\begin{eqnarray}
D_{T} &=&\lim_{\omega\to 0 }\omega_m\int\limits_{0}^{\frac
{2\pi}{\omega_m}} \int\limits_{-\infty}^{\infty}
\av{\eta_{T}^-(\omega)\eta_{T}^-(\omega')}e^{i (\omega+\omega')t}
\rd \omega' \rd t \nonumber\\
&=&\frac{1}{2}(D_m+D^-_{BA})\label{eq:Ddef}.
\end{eqnarray}
The contribution from the resonator's thermalized surroundings is $\frac{1}{2}D_m=\frac{1}{2}\gamma_m(\bar n
+\frac{1}{2})$, and the contribution from the cavity is given by,
\begin{eqnarray}
D_{BA}^\pm(z)&=& \frac{\gamma_c
g^2\Omega^2}{8}\sum\limits_n\frac{1}{|h_n|^2}\left|\frac{J_{n-1}(z)}{h_{n-1}}\pm\frac{J_{n+1}(z)}{h_{n+1}}\right|^2.
\label{eq:Dpm}
\end{eqnarray}
$D_{BA}^+$ again relates to the phase diffusion.




\begin{figure}
\center{\hspace{-0.4
cm}\includegraphics[width=4.5cm]{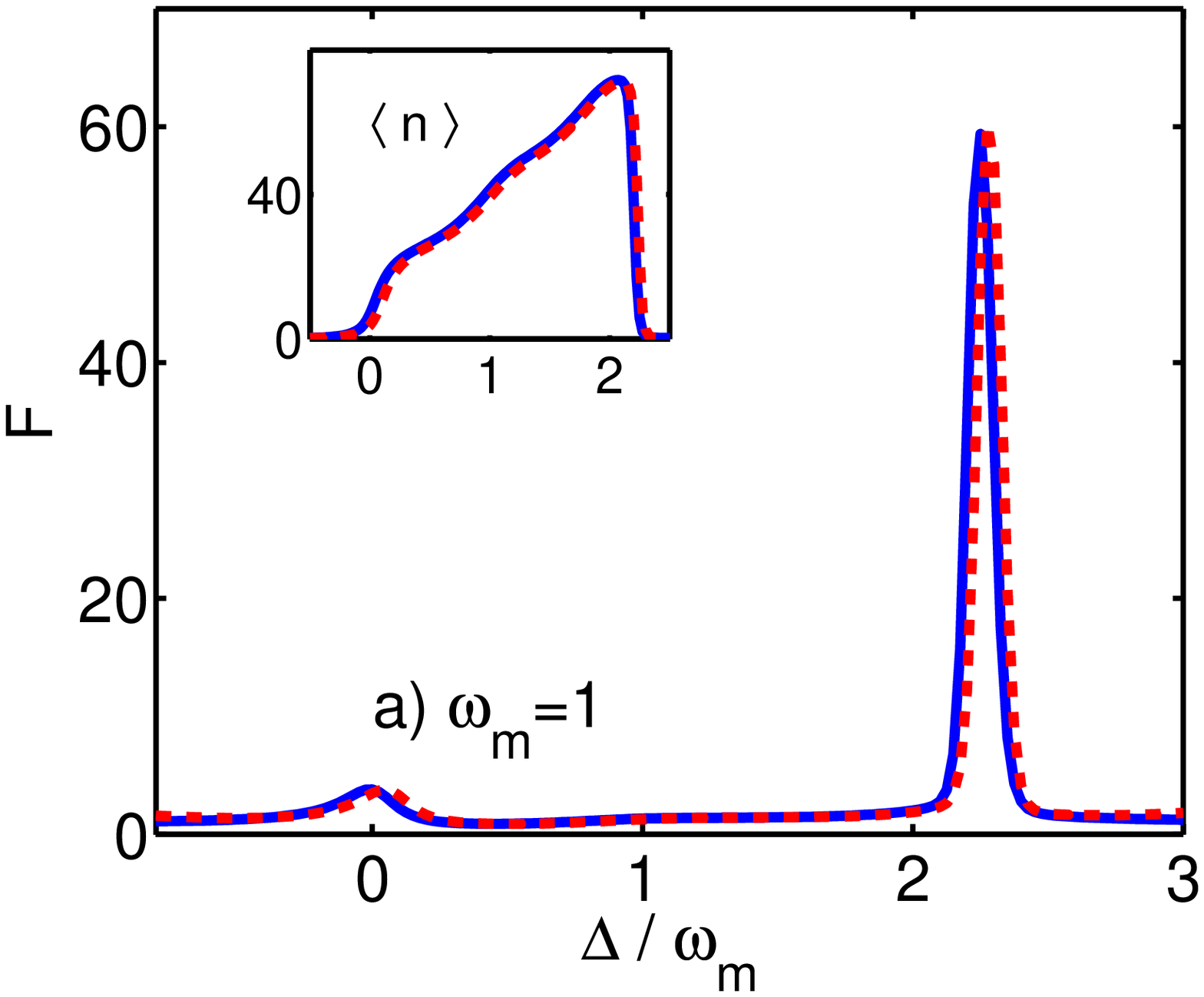}\hspace{-0.6
cm}\includegraphics[width=4.5cm]{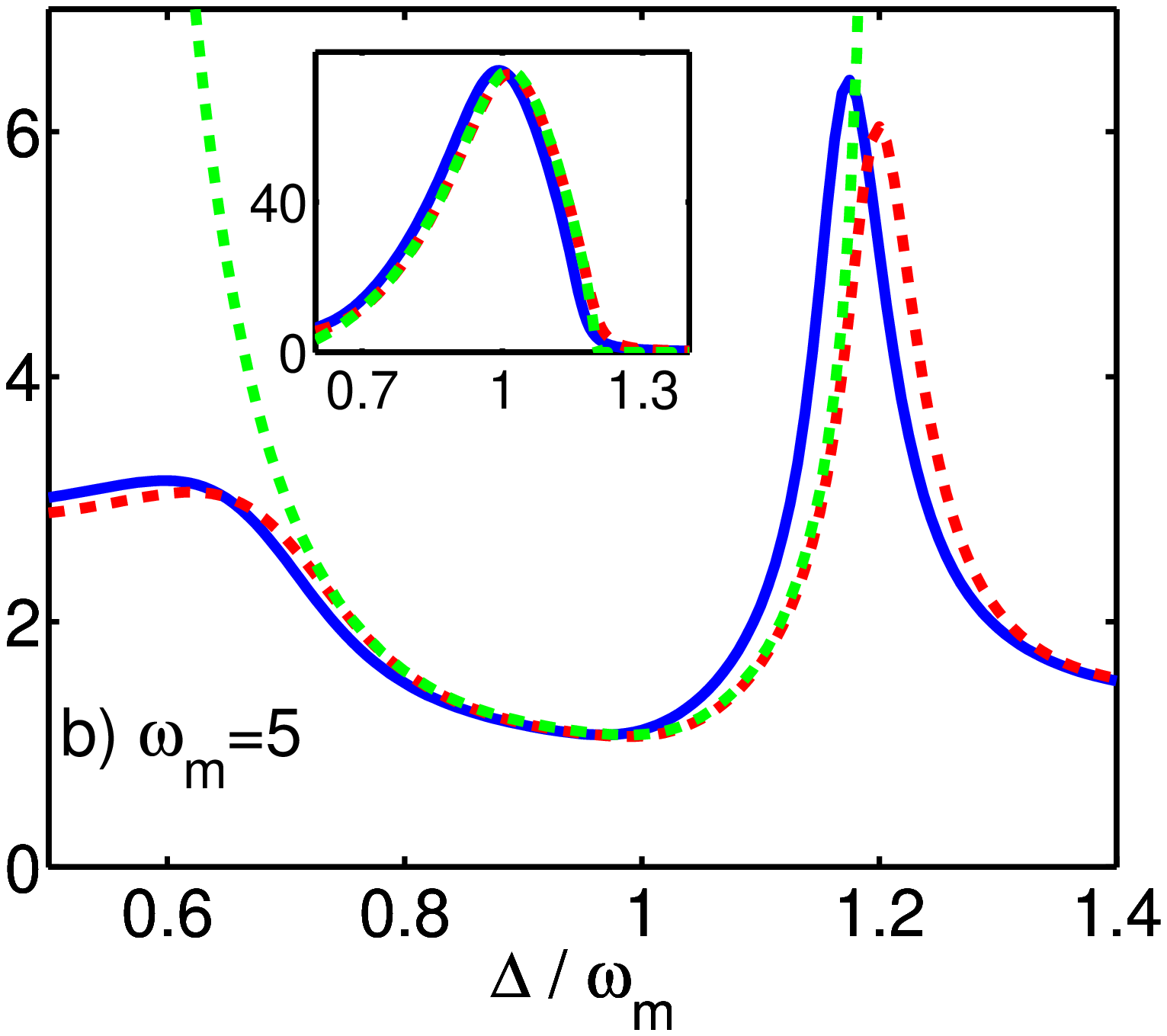} }\caption{(Color
online) Resonator Fano factor $F$ and average energy $\av{n}$
(inset) as a function of $\Delta$ calculated using $P(B)$ (dashed
red curves) and numerically (solid blue curves). In (a)
$\omega_m=1$, $g=0.4$, $\gamma_m=5\times 10^{-5}$, and in (b)
$\omega_m=5$, $g=1.5$, $\gamma_m=3\times 10^{-5}$, in each case
$\bar{n}=0$ and $\Omega=0.05$ (we adopt units such that
$\gamma_c=1$). The results obtained from the Gaussian approximation
to $P(B)$ [see text] are shown in (b) (dashed green curves).}
\label{fig:navF}
\end{figure}

The Fokker-Planck equation equivalent to Eq.\ (\ref{eq:Beom}) has a
steady-state solution $P(B)\propto  {\rm exp}(-U(B))$,
with
\begin{eqnarray}
U(B)&=& \int\limits_0^B
\frac{2B'(\gamma_m+\gamma_{BA}(B'))}{D_m+D_{BA}^-(B')} \rd B',
 \label{eq:potlb}
\end{eqnarray}
where small corrections to the drift terms due to the noise have
been neglected\,\cite{Lax,LaxIX}. This potential solution can be
used to calculate the average resonator energy, $\av{n}$ ($n=b^\dag
b$), and associated fluctuations over a wide range of parameters,
including both the good and bad cavity limits. Figure \ref{fig:navF}
shows a comparison of $\av{n}$, and the resonator Fano factor
$F=(\av{n^2}-\av{n}^2)/\av{n}$, obtained using the $P(B)$
distribution and the results of a direct numerical
solution\,\cite{LKM} of the master equation [Eq.\
(\ref{eq:master})]. The numerical calculation is performed in a
restricted number state basis using 3 states for the cavity and up
to 120 for the resonator. We also neglect elements representing
coherence between resonator states with a large separation in energy
\cite{HRA}. Representing the cavity in a basis centered on the
equivalent uncoupled ($g\rightarrow 0$) state, i.e.\
$a'=a+i\Omega/(\gamma_c/2-i\Delta)$,  allowed us to study strongly
driven cavities using only a few states (so long as the associated
\emph{variance} is not too large).

We see from Fig.\ \ref{fig:navF} that there is very good agreement
between the numerics and the calculation using  Eq.\
(\ref{eq:potlb}) when the resonator is in a limit-cycle state
(characterized in Fig. \ref{fig:navF} by a large value of $\langle
n\rangle$ and a relatively low Fano factor.) The agreement is still
quite good\,\cite{fnote} when the resonator undergoes a dynamical
transition from the limit-cycle state back to one in which it
fluctuates instead about a fixed point\,\cite{LKM,CB,HRA} (marked by
peaks in the Fano factor).
The main approximation we have made is to neglect the higher-order
derivatives (and hence higher-order correlations) by truncating the
Wigner function. The relatively strong couplings we used in order to
capture the limit-cycle dynamics numerically \cite{LKM} provide a
severe test of this approximation. The slight shift between the
analytical and numerical curves in the figures is a sign that we are
approaching the limits (in terms of coupling strengths) of the
validity of this approach \cite{foot3}.


%

In a limit cycle the resonator distribution can be approximated as a
Gaussian centered at an amplitude $B_0$, determined by the condition
$\gamma_m=-\gamma_{BA}(B_0)$, with a width given by
$\sigma^2=(D_m+D_{BA}^-(B_0))/(2B_0 \frac{\rd \gamma_{BA}}{\rd
B}|_{B_0})$,
 and a Fano factor $F\approx 4 \sigma^2$. The Gaussian approximations to $\langle n\rangle$
(given by $B_0^2-1/2$) and the Fano factor are compared with results
from numerics and using the full $P(B)$ distribution in Fig.\
\ref{fig:navF}b.

\begin{figure}
\includegraphics[width=7.5cm]{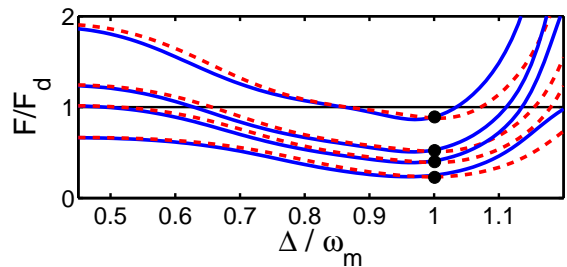} \caption{(Color online)
Analytic [Eq.\ (\ref{eq:potlb})] (dashed red curves) and numerical
(solid blue curves) calculations of the resonator Fano factor $F$ in
terms of $F_d$. The curves are (top to bottom) $\bar{n}=0,0.5,1,5$.
In each case $\omega_m=5$, $\gamma_m=3\times 10^{-5}$, $g=2$ and
$\Omega=0.05$. The dots represent the on-resonance value calculated
using Eq.\ (\ref{eq:onrF}). Note, for $\bar n=0$, $F_d=1$ and $F$
reaches a minimum $\simeq0.9$.} \label{fig:thl}
\end{figure}

The resonator Fano factor drops when it is in a well-defined
limit-cycle state. In the highly idealised case where thermal noise
is negligible ($\overline{n}\approx0$), the cavity can drive the
resonator into a non-classical sub-Poissonian state with $F<1$ (see
Fig. \ref{fig:thl}). This is a consequence of the low noise
properties of the cavity: it is well-known in the context of laser
physics that regular pumping can lead to sub-Poissonian
states\,\cite{WM}. However, for a mechanical resonator thermal noise
plays an important role and we now examine to what extent we can
think of this being suppressed in the limit-cycle states. A useful
comparison can be made between the cavity-driven resonator states
and a displaced thermal state (DTS) with the same amplitude, thermal
occupation number $\overline{n}$ and external damping $\gamma_m$.
A DTS\,\cite{DTS} is produced by harmonically driving a resonator
initially in a thermal state, increasing its energy without
introducing additional fluctuations. \trd{We choose a DTS for
comparison as it reduces to a coherent state for $\bar n \to 0$,
meaning it is both the generalization of a coherent state to finite
$\bar n$ and of a thermal state to finite amplitude.}

 The Fano factor of a DTS
with amplitude $B_0$ is $F_d=(\bar{n}(1+\bar n)+(2\bar n +1)
|B_0|^2)/(\bar n +|B_0|^2)$ and we compare this with that of the
cavity-driven resonator in Fig.\ \ref{fig:thl} for a range of
external bath temperatures. In a well-defined limit cycle  the value
of the ratio $F/F_d$ is suppressed significantly below unity and we
can think of the cavity as suppressing the thermal fluctuations.
Note that this suppression also occurs outside the good cavity limit
shown. \trd{As with the usual cavity induced cooling \cite{MCCG},
the noise suppression occurs because the low-noise cavity can
increase the friction (damping) on the resonator without adding
significantly to the diffusion. Thus, driving a mechanical resonator
via a cavity in this way produces a lower-noise state than driving
with a perfect harmonic drive.}

For $\Delta\sim \omega_m$
and in the good cavity limit,  we can use Eqs. (\ref{eq:gamma}) and
(\ref{eq:potlb}), together with the fact that $P(B)$ is almost Gaussian in a well-defined limit-cycle to obtain a simple approximate expression
for the Fano factor,
\begin{eqnarray}
F&=&\left(\bar{n}+1 +\frac{g^2B_0^2}{4\omega^2_m}\right)\left[  \frac{J_{1}(z)}{J_{1}(z)-z
J'_{1}(z)}\right]_{z=gB_0/\omega_m},\label{eq:onrF}
\end{eqnarray}
where the amplitude $B_0$ is defined by
$\gamma_m=-\gamma_{BA}(B_0)$; the predictions of this equation are
shown as dots in Fig. \ref{fig:thl}. This formula breaks down when
the cavity-resonator coupling is increased sufficiently to allow the
co-existence of more than one stable limit cycle\,\cite{MHG},
(although it correctly predicts $F$ within the second limit cycle
once the first has become unstable). Thus whilst Eq.\
(\ref{eq:onrF}) suggests that an arbitrarily small $F$ can always be
achieved, the actual minimum value achievable for a given system is
set by this expression together with the requirement that
$\gamma_m=-\gamma_{BA}(B)$ has a single (non-zero) solution.

\begin{figure}
\center{
\includegraphics[width=5.5cm]{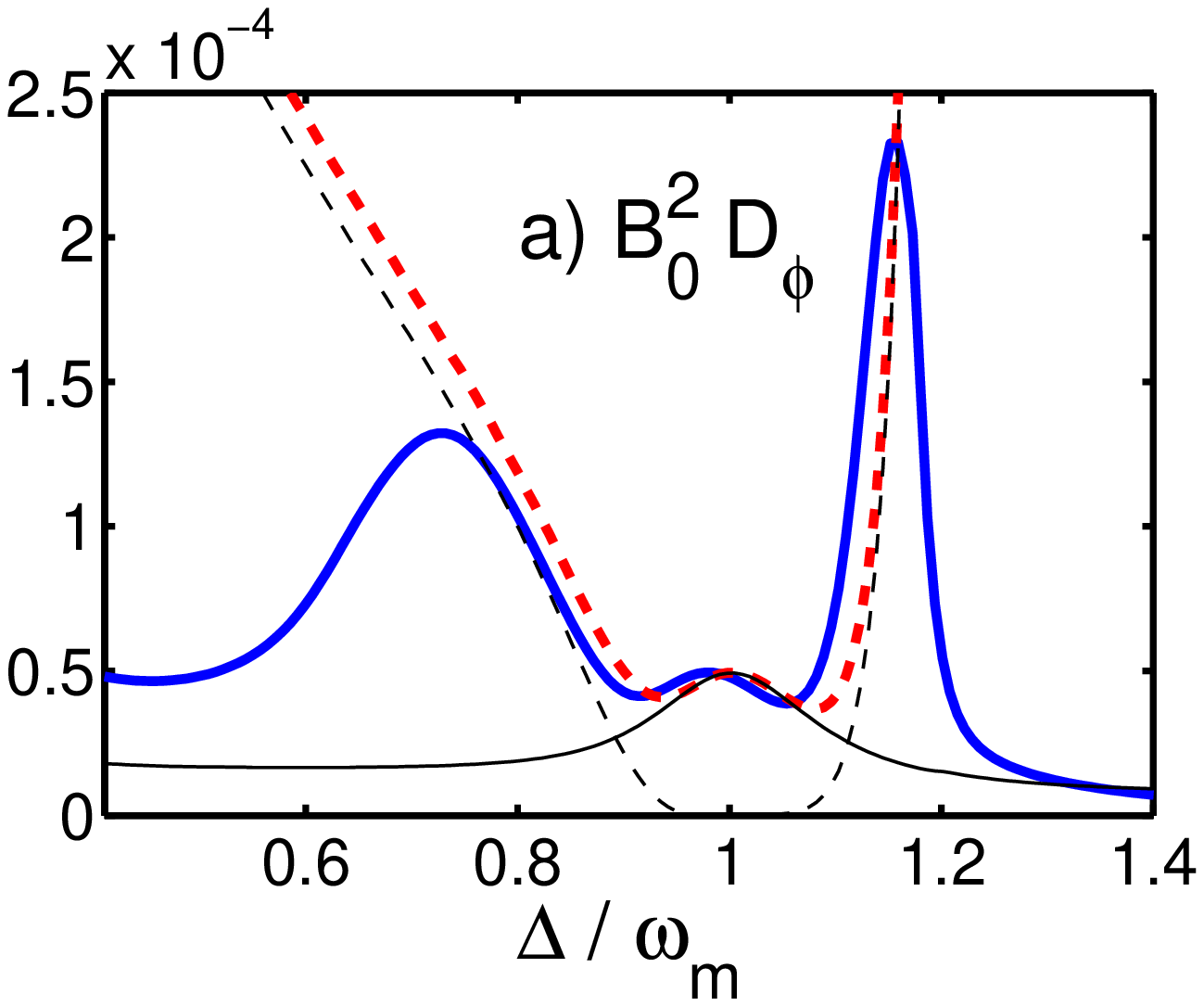}\includegraphics[width=2.75cm]{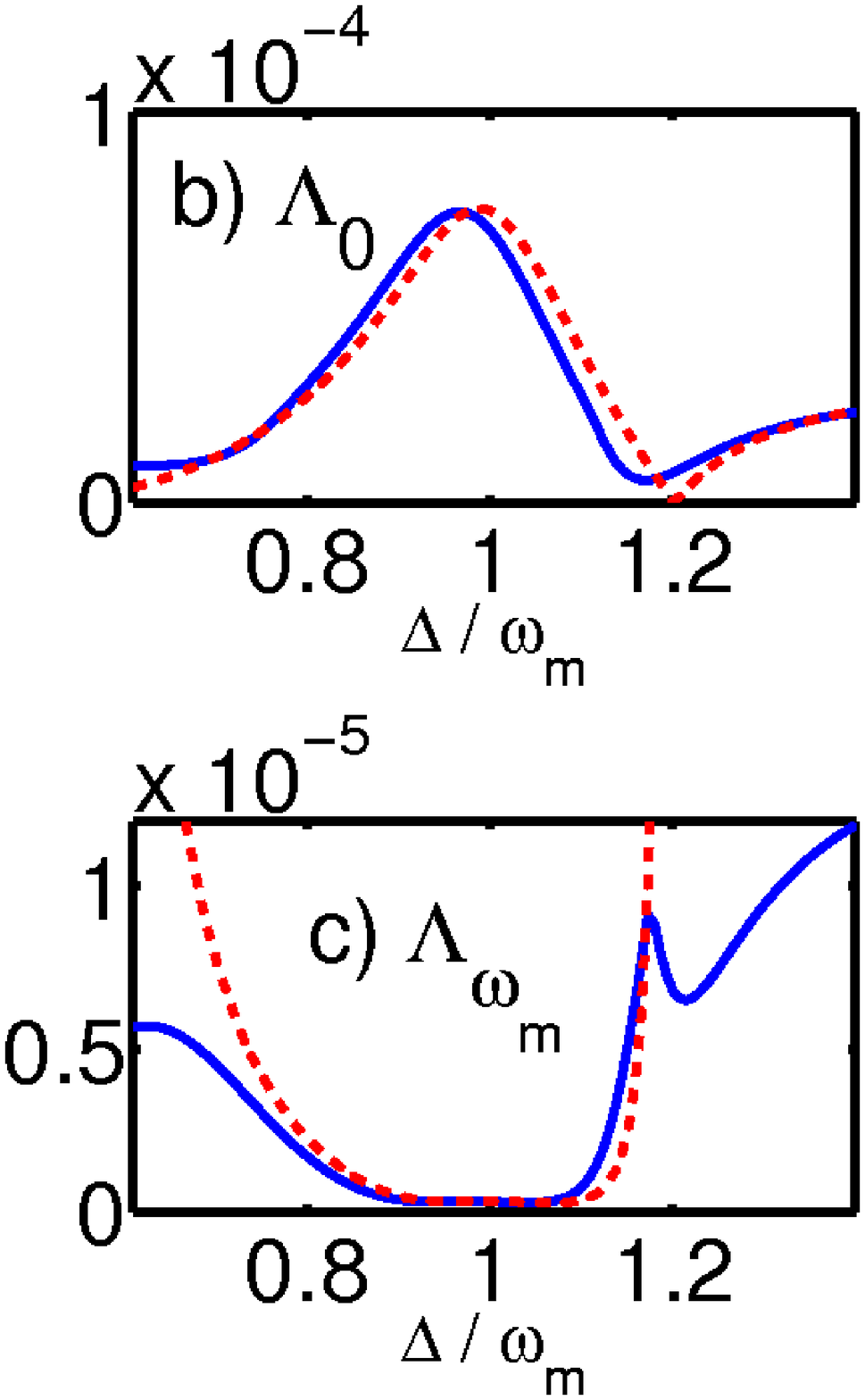}
}\caption{(Color online) a) Phase diffusion calculated numerically
(solid blue) and
 analytically (dashed red). Solid and dashed thin lines show the
 direct phase diffusion $(D_m+D_{BA}^+)/2$ and that due to the optical spring
 effect $2\delta\omega_L^2/\gamma_L^2(D_m+D_{BA}^-)$
 respectively. The insets show the linewidths $\Lambda_0$ (b) and
$\Lambda_{\omega_m}$ (c).
  Parameters are: $\omega_m=5,
 \gamma_m=3\times 10^{-5},g=1.5, \Omega=0.05, \bar n =0$.  } \label{fig:Dphi}
\end{figure}

As well as fluctuations in amplitude, the resonator also
undergoes phase diffusion which determines the linewidth in the limit-cycle state.
 In a well-defined limit-cycle, we can
write a coarse-grained equation of motion for the
phase\,\cite{Lax,CB} $\phi$, again linearizing the fluctuations,
\begin{eqnarray}
B_0\dot\phi&=& i\eta_{T}^+ +\delta \omega_L{\delta B}
\label{eq:phisol}.
\end{eqnarray}
where $\delta B$ represents the amplitude fluctuations $B-B_0$ and
$\delta\omega_L=B_0 \frac{\rd \delta \omega}{\rd B}|_{B_0}$ is the
frequency shift linearized about the limit cycle. Defining the phase
diffusion in the same way as the amplitude diffusion, Eq.\
(\ref{eq:Ddef}), we get,
\begin{eqnarray}
D_{\phi}&=& \frac{1}{2B_0^2}\left( D_m+D_{BA}^+(B_0) +
\frac{4\delta\omega_L^2}{\gamma_L^2} (D_m+D_{BA}^-(B_0) )\right)
\label{eq:Dphitot}
\end{eqnarray}
where  $\gamma_L=B_0 \frac{\rd \gamma_{BA}}{\rd B}|_{B_0}$ is the
linearized damping and cross-correlations have been neglected. Because the shift in the resonator frequency due
to the cavity (optical spring effect) is amplitude dependent, the
amplitude fluctuations can give rise to an important additional
contribution to the phase diffusion. The phase diffusion is shown in
Fig.\ \ref{fig:Dphi}  and it is clear that although the optical
spring contribution is negligible at the center of the side band
resonance (where $\delta\omega_L$ itself is negligible\,\cite{vah}),
it nevertheless becomes important on either side.

Numerical
calculations of the resonator spectrum $S(\omega)=\int dt{\rm
e}^{-i\omega t}\langle \{ b^{\dagger}(t), b(0) \}\rangle$ show
near-Lorentzian peaks around $\omega=0$ and $\omega=\omega_m$ of
width $\Lambda_0$ and $\Lambda_{\omega_m}$. These widths are
determined by the slowest dissipative timescale at each frequency
\cite{HRA,Lax}, and hence $\Lambda_0$ is given by the energy
relaxation rate $\Lambda_0=\frac{B_0}{2}\frac{\rd \gamma_{BA}}{\rd
B} \big|_{B_0}$ and the linewidth is given by the phase diffusion,
$\Lambda_{\omega_m}=\frac{D_\phi}{2}$. There is good agreement
between the numerical and analytic calculations of these quantities
(Figs. \ref{fig:Dphi}\,b,c) within the limit cycle regime (the
main difference again an effective shift in $\Delta$).


In conclusion, we have studied the amplitude and phase noise of
limit-cycle states of a mechanical resonator driven by an optical
cavity. Within a limit-cycle amplitude fluctuations are suppressed
in the sense that they can be substantially less than in a
corresponding state produced by simply applying a pure harmonic
drive, the counterpart of the cooling that occurs in the stable
regime. For low enough thermal noise the cavity generates
non-classical sub-Poissonian resonator states. However, for the
phase diffusion in the limit-cycle states the cavity noise simply
adds to the effects of thermal fluctuations and the optical spring
effect can also generate a significant contribution. The quantum
noise of the resonator is described rather well by the truncated
Wigner function approach over a range of resonator frequencies, both
within the limit-cycle states and more surprisingly within the
transition regions.


We thank T. Harvey for help with aspects of the numerics. This work
was supported by EPSRC (UK).


\begin{thebibliography}{99}
\bibitem{fnote}  The $P(B)$ description works well even in the regime defined by $g\sim\gamma_c$, where the zero-point fluctuations of the resonator are comparable to the cavity induced fluctuations\,\cite{MHG}.
\end{thebibliography}
\end{document}